\documentclass[a4paper,fleqn]{cas-dc}

\usepackage[authoryear]{natbib}
\usepackage{gensymb}
\usepackage{xpatch}
\usepackage{url}
\usepackage{tcolorbox}
\def\tsc#1{\csdef{#1}{\textsc{\lowercase{#1}}\xspace}}
\tsc{WGM}
\tsc{QE}
\tsc{EP}
\tsc{PMS}
\tsc{BEC}
\tsc{DE}

\begin{document}
\let\WriteBookmarks\relax
\def\floatpagepagefraction{1}
\def\textpagefraction{.001}

\shorttitle{Feature based sw design pattern detect}
\shortauthors{N. Nazar et~al.}

\title [mode = title]{Feature-Based Software Design Pattern Detection}                      


\author[1]{Najam Nazar}
\cormark[1]
\ead{najam.nazar@monash.edu}



\cortext[cor1]{Corresponding author.}
\author[2]{Aldeida Aleti}
\author[3]{Yaokun Zheng}


\address{Faculty of Information Technology, Monash University, Australia}

\begin{abstract}
Software design patterns are standard solutions to common problems in software design and architecture. Knowing that a particular module implements a design pattern is a shortcut to design comprehension. Manually detecting design patterns is a time consuming and challenging task, therefore, researchers have proposed automatic design pattern detection techniques. However, these techniques show low performance for certain design patterns. In this work, we introduce a design pattern detection approach, DPD$_F$ that improves the performance over the state-of-the-art by using code features with machine learning classifiers to automatically train a design pattern detector. DPD$_F$ creates a semantic representation of Java source code using the code features and the call graph, and applies the \textit{Word2Vec} algorithm on the semantic representation to construct the word-space geometric model of the Java source code. DPD$_F$ then builds a Machine Learning classifier trained on a labelled dataset and identifies software design patterns with over 80\% Precision and over 79\% Recall. Additionally, we have compared DPD$_F$ with two existing design pattern detection techniques namely \textit{FeatureMaps} \& \textit{MARPLE-DPD}. Empirical results demonstrate that our approach outperforms the state-of-the-art approaches by approximately 35\% and 15\% respectively in terms of Precision. The run-time performance also supports the practical applicability of our classifier.
\end{abstract}




\maketitle

\section{Introduction}\label{sec:intro}
Design pattern detection is an active research field and in recent years gained enormous attention by software engineering professionals~\citep{Mayvan:2017}. \citet{kuchana:2004} defines software design patterns as \textit{``recurring solutions to common problems in a given context and system of forces''}. Since their popularisation by the \textit{`Gang of Four'} (GoF)~\citep{Gamma:1995}, design patterns have been widely adopted by software professionals to improve the quality of software, and to facilitate code reuse and refactoring. Recognising that a particular software module implements a design pattern can greatly assist in program comprehension, and consequently improve software maintenance~\citep{Prechelt:2002}. Due to the increasing complexity of software projects and the differences in coding styles of software developers, detecting where in code the patterns have been implemented is not easy. 

Automatic detection of design patterns has shown to be useful in assisting software developers to quickly and correctly comprehend and maintain unfamiliar source code, ultimately leading to higher developer productivity~\citep{Walter:2016,Scanniello:2015,Gaitani:2015, Christopoulou:2012}. The majority of existing methods reverse engineer the source code to identify the design patterns~\citep{Detten:2010,De:2011} or build tools to detect design patterns in the source code e.g,~\citep{Hautamaki:2005,Moreno:2012}, and utilise code metrics e.g,~\citep{Uchiyama:2011}. Although it is relatively easy to obtain structural elements from the source code such as classes, attributes, methods etc. and transform them into graphs or other representations, they show low accuracy and fail to effectively predict the majority of design patterns~\citep{Yu:2018}. At the same time, capturing semantic (lexical) information from the source code is challenging and has not been fully attempted yet in identifying design patterns. Additionally, translating source code to natural language text has been effectively used in generating source code summaries with high accuracy~\citep{Hu:2018,Mcburney:2015,Mcburney:2016,Moreno:2013}, including our own work on identifying summary sentences from code fragments~\citep{Nazar:2016}. Based on these observations, we hypothesise that the lexical-based (basic) code features extracted from the source code will increase the accuracy of design pattern detection.

In this paper, we introduce a Feature-Based Design Pattern Detection (DPD$_F$) approach that uses source code features - both structural and lexical, and employs machine learning classifiers to predict a wide range of GOF design patterns, with higher accuracy compared to the state-of-the-art. Machine learning has been applied for DPD in the past, e.g., \citep{Fontana:2011}, however our approach is the first to employ lexical-based code features. DPD$_F$ builds a call graph and extracts 15 source code features to generate a Software Syntactic and Lexical Representation (SSLR). The SSLR provides the lexical and syntactic information of the Java files as well as the relationships between the files' classes, methods etc., in a natural language form. Using SSLR as an input, we build a word-space geometrical model of Java files by applying the \emph{Word2Vec} algorithm. We train a supervised machine learning classifier, DPD$_F$ on design patterns using the labelled dataset and a geometrical model for recognising twelve commonly used GOF software design patterns. 

To evaluate our approach, we label a corpus of 1,300 Java files extracted from a publicly available Github Java Corpus~\citep{Allamanis:2013}, which we refer to as DPD$_F$ Corpus. We use an online tool \textit{'CodeLabeller'}\footnote{\url{www.codelabeller.org}, verified on 08-10-21}~\citep{Chen:2021}, where expert raters labelled design patterns. We statistically evaluate the performance of DPD$_F$ by calculating the Precision, Recall and F1-Score measures and compare our approach to two existing software design pattern detection approaches, namely \textit{FeatureMaps} and \textit{MARPLE-DPD}. Empirical results show that our approach is effective in recognising twelve GOF software design patterns with high Precision (80\%) and low error rate (20\%). Furthermore DPD$_F$ outperforms the state-of-the-art approaches FeatureMaps \& MARPLE-DPD by 35\% and 15\% respectively in terms of Precision.

\textbf{Contributions:} In summary, this paper makes the following contributions:
\begin{itemize}
    \item We introduce a novel approach called \textit{Feature-Based Design Pattern Detector} (DPD$_F$) that uses 15 source code features to detect software design patterns.
    \item We build a large corpus i.e. DPD$_F$ Corpus consists of 1,300 Java files, which are labelled from expert software engineers using the \textit{Codelabeller} tool .
    \item We demonstrate that our approach outperforms two existing approaches with substantial margins in terms of Precision, Recall and F1-Score.
\end{itemize}

\textbf{Paper Organisation:} We discuss the preliminaries and relevant background of the related technologies, in particular, design patterns, code fragments, word space models and machine learning in Section~\ref{sec:preliminaries}. Section~\ref{sec:study_design} discusses our study design, including the research questions, how the corpora are selected and labelled. The tool used to label data, and what source code features are used and how they are selected, the application of Word2Vec for building an N-gram model and ending with a discussion of our machine learning classifier. Section~\ref{sec:results} presents results with respect to our research questions. We discuss the threats to the validity of our study in Section~\ref{sec:threats} and the related work is presented in Section~\ref{sec:relatedwork}. Finally, the conclusion is presented in Section~\ref{sec:conclusion}.

\section{Preliminaries} \label{sec:preliminaries}
In the following subsections, we briefly introduce some basic concepts that are used in this study, which are related to design patterns, code features, word-space embeddings/models and machine learning.

\subsection{Design Patterns}\label{subsec:set_of_design_patterns}

We consider the following twelve GOF design patterns: \textit{Abstract Factory}, \textit{Adapter}, \textit{Builder}, \textit{Decorator}, \textit{Factory Method}, \textit{Fa\c{c}ade}, \textit{Memento}, \textit{Observer}, \textit{Prototype}, \textit{Proxy}, \textit{Singleton} and \textit{Visitor} that are analysed by using the proposed approach. These patterns cover all three categories of GoF patterns i.e. creational, structural and behavioural patterns. Creational patterns include Builder, Abstract Factory, Factory Method, Prototype and Singleton patterns whereas, structural patterns include Adapter, Decorator, Fa\c{c}ade and Proxy patterns. We have considered three behavioural patterns for this study and that are Memento, Observer and Visitor.

\subsection{Code Features}\label{subsec:code_features}
Code features are static source code attributes that are extracted by examining the source code~\citep{McBurney:2018}, such as the size of the code elements (e.g., line of code), complexity of code such as if/else blocks, object-oriented attributes in code such as inheritance etc., and source code constructs that are uniquely identifiable names for a construct, for example class name, method name etc. \citet{Zanoni:2015} called features as code entities that are the names given to any code construct that is uniquely identifiable by its name and the name of its containers. In the object-oriented paradigm, code entities are classes, interfaces, enums, annotations, etc., methods and attributes. Previously,~\citet{Nazar:2016} used 21 textual features to identify summary lines for code fragments.

By examining aforementioned studies, we decide to use a mixture of structural and lexical object-oriented code constructs to investigate if they can be useful in identifying the design patterns from the source code. These features (as discussed in Section~\ref{subsec:features}) capture the behavioural, structural and creational aspects of the code as needed for design patterns (as discussed in Section~\ref{subsec:set_of_design_patterns}). In this paper, we follow the term features as we believe that they are more relevant to design patterns and depict the code structure and semantics better than low level entities discussed in relevant studies.

\subsection{Word Embeddings}\label{subsec:wordembeddings}
Word space models are abstract representations of the meaning of words, encoded as vectors in a high dimensional space \citep{Salton:1975}. A word vector space is constructed by counting co-occurrences of pairs of words in a text corpus, building a large square $n$-by-$n$ matrix where $n$ is the size of the vocabulary and the cell ($i, j$) contains the number of times the word $i$ has been observed in co-occurrence with the word $j$ in the corpus. The $i$-th row in a co-occurrence matrix is an $n$-dimensional vector that acts as a distributional representation of the $i$-th word in the vocabulary. 

The key to using a vector representation to compute the semantic relatedness of words lies in the \textit{Distributional Hypothesis} (DH)~\citep{Harris:1954}. The DH states that: \textit{''words that occur in the same contexts tend to have similar meaning"}, therefore allowing the computational linguist to approximate a measure of how much two words are related in their meaning by computing a numeric value from a vector space model.

The similarity between two words is geometrically measurable with any distance metric. The most widespread metric for this purpose is the cosine similarity, defined as the cosine of the angle between two vectors:
$$
S(x,y) = \frac{x \cdot y}{||x|| ||y||}
$$

Several techniques can be applied to reduce the dimensionality of the co-occurrence matrix. Latent Semantic Analysis (LSA), for instance, uses Singular Value Decomposition (SVD) to prune the less informative elements while preserving most of the topology of the vector space, and reducing the number of dimensions to the order of hundreds~\citep{Landauer:97,Mikolov:2013}. 

Recently, neural network based models have received increasing attention for their ability to compute dense, low-dimensional representations of words. To compute such representation, i.e., the word embeddings, several models rely on a huge amount of natural language texts from which a vector representation for each word is learned by a neural network. Their representations of the words are based on prediction as opposed to counting. Embedded vectors created using the predictive models such as \emph{Word2Vec} have many advantages compared to LSA~\citep{Baroni:2014}. For instance, their ability to compute dense, low-dimensional predictive representations of words. Vector spaces created on word distributional representations have been successfully proven to encode word similarity and relatedness relations~\citep{Radinsky:2011,Reisinger:2010,Ciobanu:2013,Collobert:2011}, and word embeddings have proven to be a useful feature in many natural language processing tasks~\citep{Collobert:2011,Le:2014,Santos:2014} in that they often encode semantically meaningful information of a word. 

\subsection{Machine Learning}\label{subsec:learning}

Machine learning (ML) is the study of computer programmes that learns from the data and improves automatically through experience~\citep{Mitchell:1997}. In ML the essential elements are data - structural such as text, unstructural or semi-structural such as source code, the model e.g., Support Vector Machines (SVM) or Random Forest (RF)~\citep{Cortes:1995, Ho:1995}, and the evaluation procedure e.g., cross-validation. The types of machine learning algorithms differ in their approach, the type of data they input and output, and the type of task or problem that they intend to solve. In short, there are three major types of machine learning approaches that are supervised, unsupervised and semi-supervised. Supervised learning builds a mathematical model based on the labelled data to predict future results~\citep{Christopher:2008, Bishop:2006}. Unsupervised, on the other hand, learns how systems can infer a function to describe a hidden structure from unlabelled data~\citep{Bishop:2006}. Semi-supervised falls between supervised and unsupervised approaches, since it uses both labelled and unlabelled data for training – typically a small amount of labelled data and a large amount of unlabelled data~\citep{Christopher:2008, Bishop:2006}.

In this paper, we have used supervised learning and there are two major types of supervised learning namely, classification and regression. The classification methods usually use two sets, a training set and a test set. The training set is used for learning some classifiers and requires a primary group of labelled individuals, in which the category related to each individual is obvious from its label. The test set is used to measure the efficiency of the learned classifiers and includes labelled individuals which do not participate in learning classifiers. Regression, on the other hand, allows us to predict a continuous outcome of  the variable we intend to find.

Design pattern detection can be categorised as a classification problem where classes containing the pattern can be labelled by expert raters. Thus, supervised classification learning corresponding to the ground truth that a class contains or does not contain a design pattern can be used to predict design patterns. In this study, RF and SVM based classifiers are considered from benchmark studies whereas our DPD$_F$ classifier builds a design pattern detection model that classifies given classes.

\section{Study Design}\label{sec:study_design}

In this section, we describe our research questions and justify their use in applying the overall motivation of design pattern detection from source code. Additionally, we lay out our methodology for answering research questions and the code features used in this study.

\subsection{Research Questions}\label{subsec:RQ}
This study seeks to identify design patterns through code features and \textit{Word2Vec} algorithm along with the application of supervised machine learning. In doing so, we aim to examine the relationship between code features and the design patterns. Therefore, we pose the following three Research Question (RQs):

\begin{enumerate}[1.]
    \item \textbf{RQ1}: Is DPD$_F$ effective in detecting software design patterns?
    
    The rationale behind \textit{RQ1} is to determine whether our approach identifies design patterns accurately and effectively. For this purpose, we statistically evaluate our classifiers using standard statistical measures of Precision, Recall and F1-Score.
    
    \item \textbf{RQ2}: What is the error-rate of DPD$_F$?
    
    For addressing \textit{RQ2}, we build a confusion matrix of the classifier to find how well our classifier identifies the percentage of pattern instances from the labelled data and instances are missed.
    
    \item \textbf{RQ3}: How well does DPD$_F$ perform compared to existing approaches?

    It is important to compare our approach with the existing studies to further evaluate the efficacy of our approach. To do so, we select  two studies that are FeatureMaps proposed by \citet{Thaller:2019} \& MARPLE-DPD proposed by \citet{Zanoni:2015} respectively as a benchmark studies for comparing our approach. Theses studies as well as the details of the benchmark corpus are discussed in detail in Section~\ref{subsec:methodology}
\end{enumerate}

\subsection{Methodology}\label{subsec:methodology}
Our methodology to address research questions is as follows: 

First, we collect the corpus containing Java projects. Next, we discuss how the code features are selected and applied to extract the syntactic and semantic representation (SSLR) for the corpus. After that, we apply the \textit{Word2Vec} algorithm on the SSLR  to create the geometric representation that can be read by the machine. In the end, we train supervised classifiers on the labelled corpus to predict design pattern instances from the geometric representation of Java files.

In the following subsections, we discuss these steps one by one.

\subsubsection{Data Collection}\label{subsubsec:data_col}
Existing datasets labelled with design patterns are either too small or not publicly available. We have found one publicly available corpus of Java projects called P-MART~\footnote{\url{http://www.ptidej.net/tools/designpatterns/}, verified on 08-10-21}. However, we decided to use it as a benchmark corpus. Therefore, we create a new corpus \textit{DPD$_F$-Corpus}, which we label with the respective design pattern. The aim is to increase the size of the existing corpora and make it publicly available for future researchers.

\textbf{DPD$_F$-Corpus:} We build the new corpus exclusively of design patterns from the \textit{Github Java Corpus} (GJC)~\citep{Allamanis:2013}, which is the largest publicly available corpus consisting exclusively of open source Java projects collected from GitHub. In total, the GJC contains $2,127,357$ Java files in $14,436$ Java projects. As the first step, we remove unnecessary files such as unit test cases (JUnit) or user interface files such as HTML, CSS etc. from the GJC, as these files do not normally implement design patterns. From the remainder of GJC, we select a subset of the GJC projects using the correction of finite population approach~\footnote{\url{https://www.surveysystem.com/sscalc.htm}, verified on 08-10-21} to determine the size of the final corpus that can be trained by the machine learning classifier. The final corpus has 1,300 files and we refer to it as \textit{DPD$_F$-Corpus}. To ensure that sufficient instances are available for training and testing the machine learning algorithms, we select exact 100 instances for each of the 12 design patterns (and none labels) as shown in Table~\ref{tab:no_of_instance}.


 \begin{table}[width=.9\linewidth,cols=4,pos=h]
 \centering
 \caption{The number of instances of each pattern in a labelled DPD$_F$-Corpus}
 \begin{tabular*}{\tblwidth}{@{} LLLL@{} }
 \toprule
 \textbf{Patterns} & \textbf{\#} &\textbf{Patterns} & \textbf{\#} \\
 \midrule
    Abstract Factory  & 100         & None              & 100          \\
    Adapter           & 100         & Observer          & 100          \\
    Builder           & 100         & Prototype         & 100          \\
    Decorator         & 100         & Proxy             & 100          \\
    Factory Method    & 100         & Singleton         & 100          \\
    Fa\c{c}ade        & 100         & Visitor           & 100          \\
    Memento           & 100         &                   &              \\
 \bottomrule
 \end{tabular*}
 \label{tab:no_of_instance}
 \end{table}

\textbf{Benchmark Corpus:} In addition to the  \textit{DPD$_F$} corpus, we employ an existing dataset - P-MART used by benchmark studies~\citep{Zanoni:2015,Thaller:2019}, which contains 4,242 files from 9 projects, which are QuickUML, Lexi, JRefactory, Netbeans, JUnit, JHotDraw, MapperXML, Apache Nutch and PMD. On exploring the P-MART we found that it contains uneven number of design patterns and requires surgery to fit for our purpose. The P-Mart corpus contains 1,039 files that are labelled as design patterns we plan to identify in this study. Table~\ref{tab:P-MART} shows the distribution of 12 design patterns in the P-MART Corpus. On exploring the P-MART we found that it contains an uneven number of design patterns as shown in Table~\ref{tab:P-MART}. 

\begin{table}[width=.9\linewidth,cols=4,pos=h]
\centering
\caption{The number of instances of each pattern in a P-MART corpus}
\begin{tabular*}{\tblwidth}{@{} LLLL@{}} 
\toprule
\textbf{Patterns} & \textbf{\#} & \textbf{Patterns} & \textbf{\#} \\ 
\midrule
Abstract Factory  & 241         & Observer          & 137          \\
Adapter           & 241         & Memento           & 15           \\
Builder           & 43          & Prototype         & 32           \\
Decorator         & 63          & Proxy             & 3            \\
Factory Method    & 102         & Singleton         & 13           \\
Fa\c{c}ade        & 11          & Visitor           & 139          \\
\bottomrule
\end{tabular*}
\label{tab:P-MART}
\end{table}

\subsubsection{Data Labelling}\label{subsubsec:labelling} 
We use an online tool \textit{CodeLabeller} that is built using Node and Angular JavaScript languages to label both corpora via crowd knowledge. Each file in the corpora is labelled by at least three annotators, who have at least 2 years of programming experience in Java programming language, applied design patterns in the projects they have worked on, and taught software engineering and related units at the university level. We have asked the faculty members, postgraduate and doctoral students majoring in Software Engineering and Computer Science at the Faculty of IT, Monash University to label the corpora through our online tool. Some files are labelled as \textit{'None'}, which means that they do not implement or contain any of the design patterns. We added None files in our corpus solely to check how well our classifier identifies files that do not contain a design pattern. Furthermore, if a pattern is implemented in more than one file, all the files that are part of the pattern are labelled with the respective pattern. For example, a Fa\c{c}ade pattern has two files: interface and implementation, both of which are labelled as Fa\c{c}ade.

\textbf{Raters Agreement:} Labelling is a subjective process and there is a possibility that a file is labelled differently by different annotators. Following a similar approach to \citep{Nazar:2016}, we perform a Cohen's Kappa Test~\citep{Cohen:1960} to measure the level of agreement among annotators. For the DPD$_F$ corpus, the kappa K-value is \textit{0.74}, showing a medium to high level of agreement among raters~\citep{Cohen:1960,Carletta:1996}. We have not calculated the Kappa value for the benchmark corpus.

\subsection{Feature Extraction}\label{subsec:features}
Feature extraction deals with the inherent complexity of programming languages by extracting high-level concepts that in later stages can be used to successfully find pattern instances. Structurally, design patterns describe classes and their loose arrangements and communication paths. Consequently, extracted features ideally capture these arrangements and their relationships to improve reasoning in later stages. As discussed in Section~\ref{subsec:code_features} the source code features are high level code constructs that are used to extract information from the source code. Following a similar approach with~\citet{McBurney:2018} \& \citet{Nazar:2016}, we have selected 15 features that are based on the syntactic and semantic (the linguistic meanings) constructs of Java source code. 

As listed in Table~\ref{tab:features}, these features are related to \textit{class labels},\textit{ method names}, \textit{modifiers}, \textit{attributes}, \textit{n-grams}, \textit{method return type}, \textit{number of parameters}, \textit{the number of lines in a method}, \textit{incoming methods} and \textit{outgoing methods}. We may informally categorise these features into two major groups that are class level features (4 features in total), and method level features (11 features in total).

\subsubsection{Class-Level Features}\label{subsubsec:Class-levelfeatures}
A class is a user-defined blueprint or prototype from which objects are created and in the Java language, the class name begins with a capital letter so the first code feature we select is a \textit{`ClassName'} i.e., the name of the Java class. The second feature - feature 2 - is about the access modifiers in Java language i.e., public, private, protected and default. Next two features are features 3 and 4 - implements \& extends are Java keywords - and related to the inheritance principle of object-oriented languages. Inheritance is key to many patterns such as observer, abstract factory etc.; therefore, it is important to capture it. In Java language, there are two keywords for inheritance i.e. \textit{extends} when a class inherits a class and \textit{implements} when a class inherits an interface. An interface can extend another interface in the same way that a class can extend another class; therefore, the extends keyword can also be used to extend an interface.

\subsubsection{Method-Level Features}\label{subsubsec:method}

A Java  method is a collection of statements that are grouped together to perform an operation. In general, a method in Java contains a name, which begins with a small letter in a camelCase format, one or more parameters (sometimes known as attributes) enclosed in parentheses and a return type - these are features 5, 6 \& 7 respectively in our case. The Feature 8 measures properties of the statements inside the method's body. These statements can be local attributes, conditional statements such as if, else, switch, or an assignment statement etc. Feature 9 measures the number of variables or attributes in the method - some methods may have local scoped variables while some may not. Feature 10 is about the number of methods called within a method, whereas feature 11 calculates the number of lines within a method's body. `MethodIncomingMethod' - feature 12 - is about the number of methods a given method calls. For example, Method A calls Methods B, C and D then the value for `MethodIncomingMethod' is 3. The difference between feature 10 and 12 is that feature 10 numbers all the methods which are called within a method that can be the methods from the same class or from other classes. Whereas, feature 12 only provides the number of methods that belong to external classes. For example a method A in class A has called method B from class B and Method C from class A then the value for `MethodIncomingMethod' will be 1. Feature 13 is related to method incoming function as it illustrates the names of methods that call Method A. Similarly, features 14 and 15 are related to each other as they mention the number of methods a given method - say method A - has been called from and their names.

\begin{table}[htb]
    \caption{15 source code features \&  their descriptions}
    \begin{tabular*}{\tblwidth}{@{} LLp{4.2cm} @{} }
    \toprule
        \textbf{No.} & \textbf{Features} & \textbf{Description}\\
        \midrule
        1 & ClassName & Name of Java Class. \\
        2 & ClassModifiers & Public, Protected, Private Keywords etc.\\
        3 & ClassImplements & A binary feature (0/1) if a class implements an interface.\\
        4 & ClassExtends & A binary feature (0/1) if a class extends another class.\\
        5 & MethodName & Method name in a class.\\
        6 & MethodParam & Method parameters (arguments).\\
        7 & MethodReturnType & A method that returns something (void, int etc.) or a method having a return keyword.\\
        8 & MethodBodyLineType & Type of code in a method's body e.g. assignment statement, condition statements etc.\\
        9 & MethodNumVariables & Number of variables/attributes in a method.\\
        10 & MethodNumMethods & Number of method calls in a class.\\
        11 & MethodNumLine & Number of lines in a method. \\
        12 & MethodIncomingMethod & Number of methods a method calls.\\
        13 & MethodIncomingName & Name of methods a method calls.\\
        14 & MethodOutgoingMethod & Number of outgoing methods.\\
        15 & MethodOutgoingName & Name of outgoing methods.\\
    \bottomrule
    \end{tabular*}
    \label{tab:features}
\end{table}

\subsection{Feature-Based Design Pattern Detector}\label{subsec:feature_based_pattern}
\begin{figure*}
	\centerline{\includegraphics[width=\linewidth]{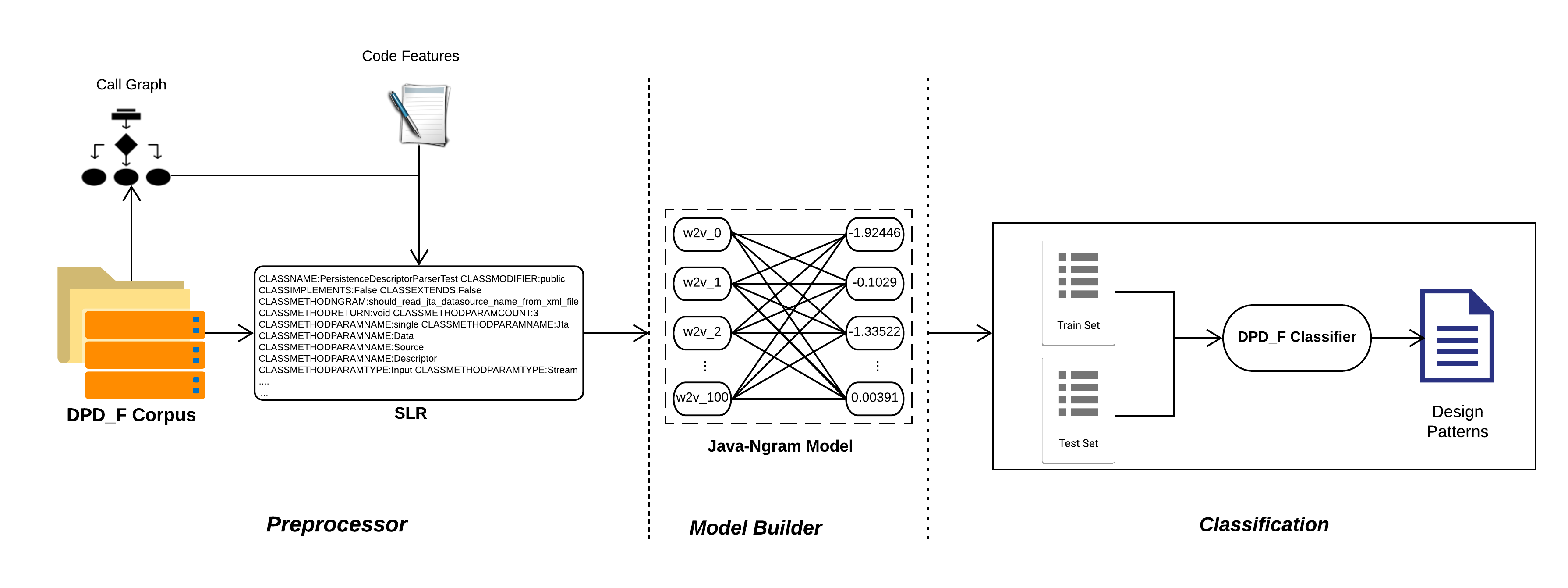}}
	\caption{Design Pattern Detection with Features (DPD$_F$).}
	\label{fig:approach}
\end{figure*}

As shown in figure~\ref{fig:approach}, the DPD$_{F}$ approach has three main steps namely \textit{Preprocessing}, \textit{Model Building} and \textit{Machine Learning Classification}. The Preprocessing step includes building a call graph and generating the SSLR representation by parsing the Java source code with code features. Once the code is parsed into natural language representation (SSLR), we apply the Word2Vec algorithm on SSLR to generate a Java Embedded Model (JEM) in the model building step. The JEM is used as an input to a supervised machine learning classifier trained on design patterns to predict design patterns. In the subsequent sections, we discuss these steps in more detail.

\subsubsection{Preprocessing}\label{subsubsec:preprocess}
Translation between source code and natural language is challenging due to the structure of the source code. One simple way to model source code is to just view it as plain text. However, in such a way, the structure and semantic information will be omitted, which may cause inaccuracies in the generated natural language text~\citep{hu:2020}. The structural information is extracted using Static Call Graphs (SCG), which encode how methods and functions call each others~\citep{musco:2017} and build a graph (edges and nodes) of flow of code to show the relationship between classes and methods in the source code. 

\begin{figure}[htbp]
    \centering
    \includegraphics[width=\linewidth]{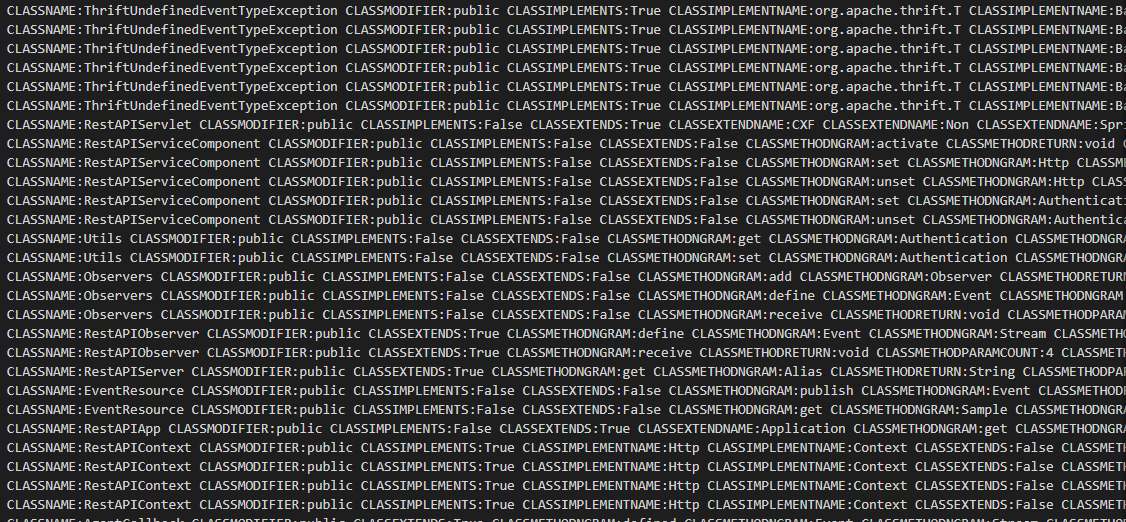}
    \caption{An example (subset) of the SSLR file.}
    \label{fig:verbose}
\end{figure}

Furthermore, building a call graph is useful as we need to encode the caller-callee classes and methods in the corpus which is the basis of generating the natural language text from the source code. Simply having a code hierarchy may not provide the complete (or enough) information we need for building the SSLR file. Based on these arguments, SCG will be more helpful in extracting the structural, behavioural and creational aspects of design pattern compared to AST. Thus, we generate a natural language representation of the source code i.e. SSLR by parsing the source code with code features and the SCG to encode the relationships between the code elements. The code elements are classes, methods, interfaces, etc. in the \textit{DPD$_F$-Corpus}. The figure~\ref{fig:verbose} illustrates a small subset (example) of SSLR file generated for the Observers.java file from the Platform project that is the part of the labelled corpus.

\textit{Implementation:} We have implemented a parser in the Python programming language that parses Java files using the \textit{Plyj}\footnote{\url{https://github.com/musiKk/plyj}, verified on 08-10-21} library. We have written our own callgraph generator and feature extractor to build the SSLR representation of the corpus. Both files parse the input corpus and generate the SSLR representation. \textit{Plyj} is a Java 7 parser written in Python using Ply\footnote{\url{https://github.com/dabeaz/ply}, verified on 08-10-21} and as the \textit{'DPD$_F$ corpus’} contains files from the projects that were written before 2013, thus, the files are written prior to the release of Java 8 (pre- Java8) - using \textit{Plyj} fulfils our requirement.

\subsubsection{Model Building}\label{subsec:model} 
We build a Java embedded n-gram model representation, by applying the \textit{Word2Vec} algorithm~\citep{Baroni:2014} on the SSLR representation of the DPD$_F$ corpus generated in the preprocessing step. Word2Vec is a method to build distributed representation of words based on their contexts that works in two alternative fashions: that are the \textit{Continuous Bag-of-Words} (CBOW) model, which predicts the target word from source context words, and the Skip-Gram model, which does the inverse and predicts source context words from the target words.

The CBOW model (also called the CBOW architecture or Vector Model) is meant for learning relationships between pairs of words. In the CBOW model, context is represented by multiple words for a given target word~\citep{Mitchell:2008}. For example, we could use \textit{`cat'} and \textit{`tree'} as context words and \textit{`climbed'} as the target word. The Skip-gram model architecture, on the other hand, predicts the source context words (surrounding words) given a target word (the centre word).

Hence, we create a Word Space Model of n-grams extracted from the content of the Java source code files using the source code features. We treat these files as if it was a natural language document, extracting the n-grams by segmenting the names of the most salient elements i.e., classes and methods etc., and consider them as words in the document. We then run the Word2Vec algorithm on the dataset to construct a high-dimensional representation of these n-grams, i.e., each n-gram is paired with a high-dimensional vector in a continuous dense space. The vectors representing the n-grams occurring in a Java class are composed by averaging them and the resulting vector is concatenated to the features extracted from the Java class with previous methods.

We use the CBOW architecture, which is meant for learning relationships between pairs of words, i.e., classes and methods mainly as in our case. We set a matrix of size 100 to build a 100-dimensional embedding model which is trained with \textit{Word2Vec} that results in the vector representation of each ngram in our collection; the vector representation of a Java file will therefore be a function of the vector representations of its n-grams. Inspired by~\citep{Mitchell:2008}, we produce embeddings for the Java files as a uniform linear combination of the embeddings of its constituent n-grams using Word2Vec.

In the 100-dimensional vector, each instance is associated with its project id, the class name, a 100-dimensional feature vector derived from a word embedding model built from the n-grams in the Java file, and a design pattern label where the design pattern is the target label to predict. These word embeddings provide a compact yet expressive feature representation for Java classes in a project source code. Despite being a programming language, and therefore a formal one, as opposed to natural language, several elements of the source code are natural language, including its name, the names of its methods and variables,  and the comments etc.

\textit{Implementation:} We use the \textit{Word2vec} implementation provided by \citep{Rehurek:2010} from the \textit{Python Gensim Library}\footnote{\url{https://radimrehurek.com/gensim_3.8.3/index.html}, verified on 08-10-21} for generating Java n-gram model. The generated SSLR representation created in the preprocessing step is passed as an input and a Java Embedded Model is generated as a result of the model generation step.

\subsubsection{Machine Classification}\label{subsec:classification}
By parsing the files in the labelled corpus, we are able to build a large dataset of these files paired with bags of n-grams relevant to each Java file. This structure is comparable to a natural language corpus, by drawing parallels between a Java source code (n-grams) and a text document. The augmented feature representation in Section~\ref{subsec:model} is used to train our DPD$_F$ classifier. 

The DPD$_F$ classifier is an ensemble classifier that uses randomised decision trees as a base estimator. It implements a meta-estimator that fits a number of randomised decision trees on various subsamples of a dataset and uses ensembling to improve the predictive accuracy and control over-fitting. As we are dealing with the multi-class classification, we use SAMME-R as a boosting algorithm~\citep{Hastie:2009}.

\begin{table}[]
\begin{tabular}{|l|l|}
\hline

\end{tabular}
\end{table}

\begin{table}[width=.9\linewidth,cols=4,pos=h]
\centering
\caption{The DPD$_F$ classifier's learning parameters}
\begin{tabular*}{\tblwidth}{@{}LL@{}} 
\toprule
\textbf{Parameters} & \textbf{Values} \\ 
\midrule
Base Estimator      & Random Forest   \\
No of Estimations   & 100             \\
Learning Rate		& 1				\\
Algorithm           & SAMME.R         \\
\bottomrule
\end{tabular*}
\label{tab:algorithm}
\end{table}

\textit{Implementation:} we use Python's \textit{Scikit-learn}~\citep{scikit-learn} library for building our classifier and measuring the efficacy of the classification using standard statistical measures (discussed in Section~\ref{subsec:criteria}). Table~\ref{tab:algorithm} provides a brief summary of our classifier's learning parameters.

\begin{figure*}[htbp]
\centering
   \includegraphics[width=0.8\linewidth]{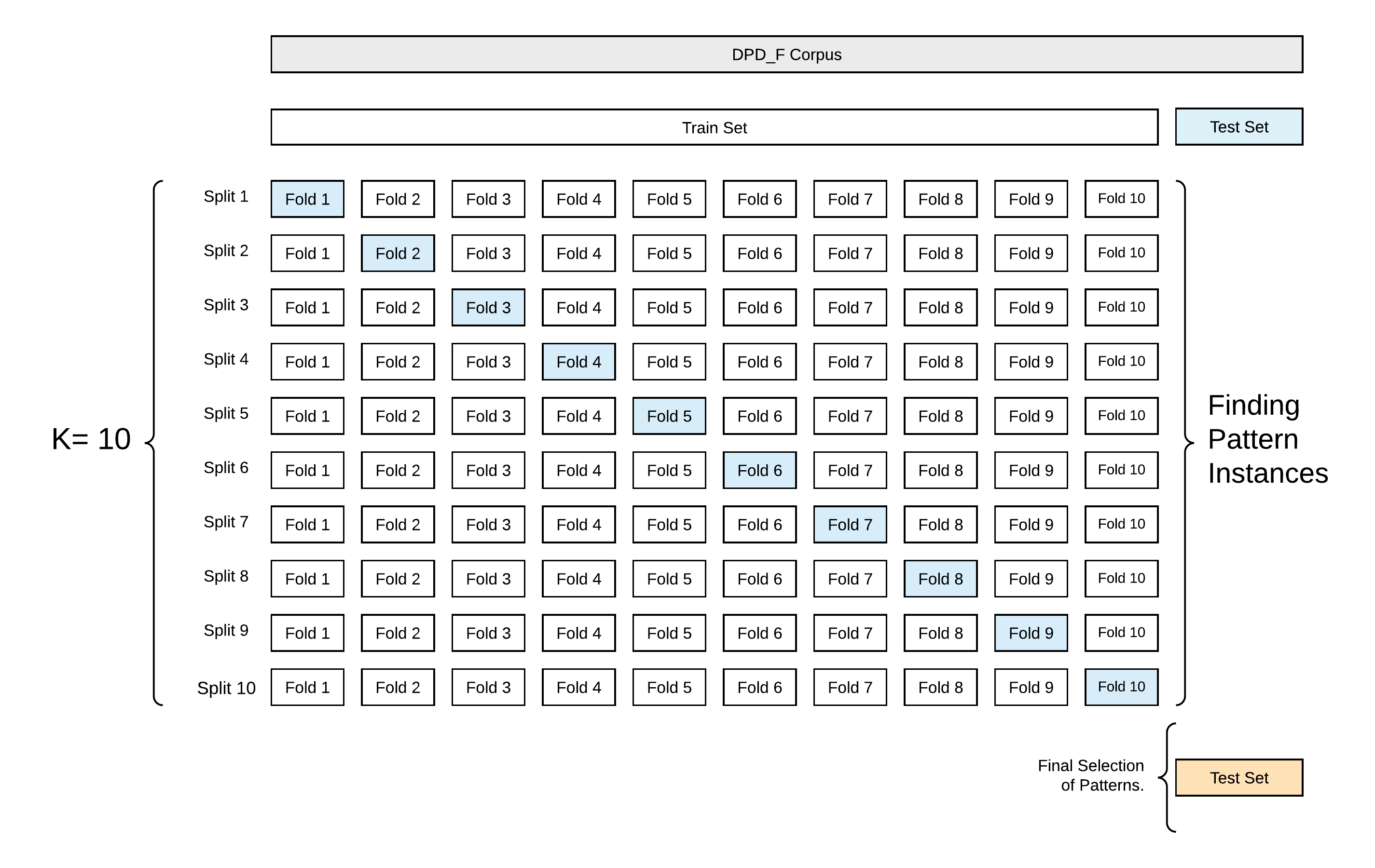}
    \caption{Stratified K-Fold Cross-Validation}
   \label{fig:kcv}
\end{figure*}

\textbf{Cross-Validation:} Cross-Validation is a resampling procedure that is used to evaluate the machine learning models on a data sample. We have used the \textit{Stratified K-Fold Cross-Validation} (SK-Fold CV) procedure for evaluating our machine learning model. The K-Fold Cross Validation is a type of Cross-Validation that involves randomly dividing the set of observations into K groups, or folds of approximately equal size~\citep{James:2013}. Normally, the first fold is treated as a validation set, and the method is fit on the remaining K minus 1 folds. A value of K=10 is the recommended value for k-fold cross-validation in the field of applied machine learning~\citep{kuhn:2013}, thus, we have used 10 folds per cross validation to validate our machine learning model. As there is a variation in sample size for each instance of design pattern in our corpus, it is important that each fold contains the same percentage of each design pattern instance to have fair prediction. Stratification is a variation of traditional K-Fold CV which is defined by \citep{kuhn:2013} as \textit{'the splitting of data into folds ensuring that each fold has the same proportion of observations with a given categorical value'} that is pattern instances for our model. Figure~\ref{fig:kcv} illustrates our K-Fold Stratification procedure using 90/10 train test splits on the DPD$_F$ corpus.

\textit{Implementation:} Our classifier used the sklearn StratifiedKFold~\footnote{\url{https://scikit-learn.org/stable/modules/generated/sklearn.model_selection.StratifiedKFold.html}, verified on 08-10-21} implementation of cross-validation.

\section{Results and Evaluations}\label{sec:results}
This section presents the results of our study where our answers for each research question are presented, evaluated and supported by our data and interpretation.

\subsection{Evaluation Criteria}\label{subsec:criteria}
We report the Precision, Recall, and F1-Score, which are the standard measures to statistically evaluate the efficacy of classifiers.

\textbf{Precision:} Precision \textit{(P)} is defined as the fraction of instances of a classification that are correct, calculated as in equation~\ref{eq:precision}:

\begin{equation}\label{eq:precision}
	P = \frac{TP}{TP+FP}
\end{equation}

Where \textit{TP} stands for true positives and \textit{FP} stands for false positives.

\textbf{Recall:} Recall \textit{(R)} is defined as the proportion of actual instances of a classification that are classified as such by the classifier as shown in the equation~\ref{eq:rec}. FN in the equations stands for false negatives.

\begin{equation}\label{eq:rec}
	R = \frac{TP}{TP+FN}
\end{equation}

\textbf{F1-Score:} The F1-Score \textit{(F1-S)} is measured as the harmonic mean of P and R and it is computed as follows:

\begin{equation}\label{eq:fscore}
	F1-S =   2* \frac{P * R}{P + R}
\end{equation}

\subsection*{RQ1: Is DPD$_F$ effective in detecting software design patterns?}

\begin{figure*}[htbp]
 \centering
   \includegraphics[scale=0.6]{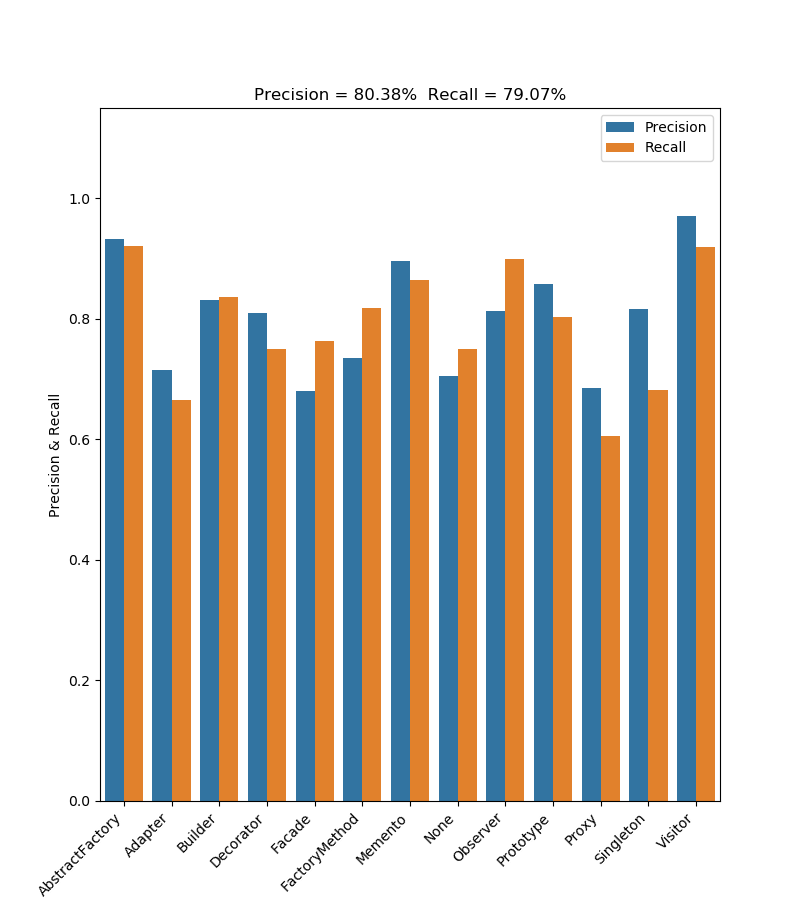}
     \caption{Precision \& Recall for the DPD$_F$ classifier.}
   \label{fig:graph}
 \end{figure*}
Since the initialisation of the classifier is random, the results slightly vary at each run, although the difference is insignificant and there are no striking differences in performance, and the overall performance is very similar for each iteration. Therefore, we calculate the weighted average of each turn to determine the final Precision, Recall and F1-Score for each label. Using the DPD$_F$ corpus, the DPD$_F$ classifier is able to predict most of the labels accurately, reaching a Precision of more than 80+\% and Recall of 79+\%. Figure~\ref{fig:graph} and Table~\ref{tab:resultsour} shows the results obtained from the DPD$_F$ classifier broken down into labels.

\begin{table}[width=.9\linewidth,cols=4,pos=h]
\caption{Precision, Recall and F-Score values for every label returned by the DPD$_F$ classifier.}
\begin{tabular}{@{}lrrr@{}}
\toprule
\textbf{Classifier} & \multicolumn{3}{c}{\textbf{DPD$_F$}}                             \\ \midrule
\textit{Design Patterns}   & \textit{Precision (\%)} & \textit{Recall (\%)} & \textit{F1-Score (\%)} \\ \midrule
\textit{Abstract Factory}   & 93.27 & 92.08 & 92.46 \\
\textit{Adapter}            & 71.56 & 66.55 & 68.41 \\
\textit{Builder}            & 83.21 & 83.66 & 82.36 \\
\textit{Decorator}          & 80.99 & 75  & 77.34 \\
\textit{Fa\c{c}ade}         & 68 & 76.27 & 71.06   \\
\textit{Factory Method}     & 89 & 83.88 & 85.79 \\
\textit{Memento}            & 89.66 & 86.44 & 87.45 \\
\textit{Observer}           & 81.26 & 90 & 85.06 \\
\textit{Prototype}          & 85.75 & 80.33 & 82.59 \\
\textit{Proxy}              & 68.51 & 60.55 & 62.86 \\
\textit{Singleton}          & 81.6 & 68.22 & 72.62 \\
\textit{Visitor}            & 97.07 & 91.88 & 93.39\\
\textit{None}               & 70.44 & 75 & 71.91 \\ \bottomrule
\end{tabular}
\label{tab:resultsour}
\end{table}

As evident from the results the most of the patterns are easily recognised by DPD$_F$ classifier, with \textit{Visitor} having the highest Precision of approximately 97\% and (interestingly) \textit{Fa\c{c}ade} with the lowest Precision of 68\%. We believe the comparatively low Precision score for Facade and Proxy is due to their complex structure and usage in different context, e.g. \textit{Proxy} can be mislead with the network proxies. Other patterns such as \textit{Abstract Factory}, \textit{Factory Method}, \textit{Memento} and \textit{Builder} are very well recognised by DPD$_F$ with over 80\% Precision. In conclusion, our DPD$_F$ classifier detects all patterns on average with over 80\% of Precision and 79\% of Recall.

\subsection*{RQ2: What is the error rate of DPD$_F$?}

It is important to find that if the classifier misses some instances while training or predicting the instances or misidentifies these instances. We calculate the misclassification rate or error rate for this purpose. 

\begin{figure*}[!ht]
  \centering
    \includegraphics[scale=1]{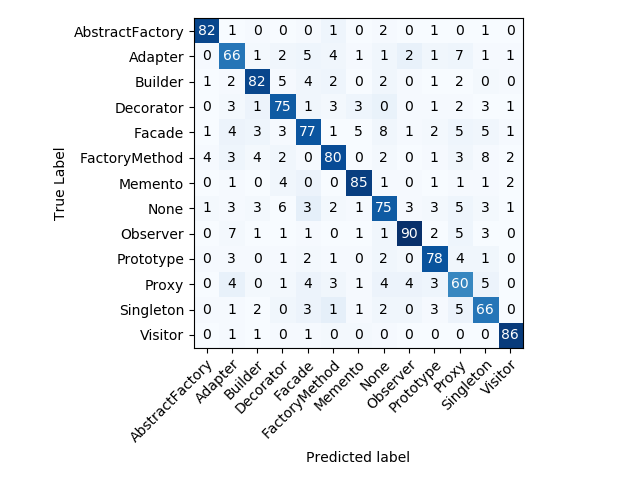}
    \caption{Confusion Matrix of the DPD$_F$ classifier.}
   \label{fig:confusionM}
  \end{figure*}

Error rate is the ratio of how often the classifier is wrong or predicts the instances incorrectly. We compute the \textit{confusion matrix} for DPD$_F$, which is a table that is often used to describe the performance of a classifier on a set of test data for which the true values are known. The results in Figure~\ref{fig:confusionM} show that DPD$_F$ performs very well in detecting the absence of a design pattern, i.e., a 'none' - 75 out of 100 instances are true values. Similarly, 86 out of 100 instances for Visitor, 90 out of 100 instances for Observer, 85 out of 100 for Memento, 82 out of 100 for Builder and Abstract Factory are truly predicted by DFD$_F$. Looking at the confusion matrix, some of the instances are wrongly identified or missed by the classifier. For example, seven instances of Observer pattern are identified as Adapter pattern. A fair number of instances for Singleton, Proxy and Adapter are missed by the classifier and the truly predicted instances for these patterns are slightly lower than the other patterns. Figure~\ref{fig:confusionM} shows the confusion matrix for DPD$_F$ classifier. Overall, the misclassification or the error rate is less than 20\% and DPD$_F$ has correctly identified 80\% of the instances of the all design patterns with high Precision and Recall.

\subsection*{RQ3: How well does DPD$_F$ perform compared to existing approaches?}\label{subsec;RQ3}

It is observed that due to the lack of publicly available standard benchmark, the evaluation and validation of the accuracy of classifiers is difficult. Either existing approaches do not share the corpus or the source code is publicly unavailable to replicate the results.  

\textbf{Selection Criteria:} Under these limitations, we select two studies from the literature that are relevant to our study. The relevancy measure is that they have either utilised code features or machine learning models to predict design patterns or a combination of both. Though they have not shared their implementation for replicating the results, we select two approaches  which are clearly described, which helps us to replicate the results for comparison purposes.

\textbf{Benchmark DP detection approaches:} Based on the selection criteria discussed above we compare our study with existing approaches in design pattern detection. The benchmark approaches are developed by ~\citet{Thaller:2019} and ~\citet{Fontana:2011}. These studies have utilised some level of code features and machine learning classifiers to identify design patterns. Since the source code and the (part of)\footnote{The DPExample is not publicly available} corpus are not publicly available, we reproduce and replicate the results using the information provided in the existing studies to compare with our study.

\begin{table*}
\caption{Comparison of DPD$_F$ with the state-of-the-art. The best results are presented in bold font. Precision (P), Recall (R) and F1-Score (F1-S) values of the benchmark approaches are generated for both benchmark (Labelled P-MART) and our DFD$_F$ corpus. The P-MART corpus contained some of the patterns and not all so the results are tested for the patterns mentioned in the P-MART corpus only.}
\begin{tabular}{@{}ll|rrr|rrr|rrr@{}}
\toprule
\textbf{Corpora} &
  \textbf{Design Patterns} &
  \multicolumn{3}{c|}{\textbf{FeatureMap}} &
  \multicolumn{3}{c|}{\textbf{MARPLE-DPD}} &
  \multicolumn{3}{c}{\textbf{DPD$_F$}} \\ \midrule
 &                    & P (\%) & R (\%) & F1-S (\%) & P (\%) & R (\%) & F1-S (\%) & P (\%) & R (\%) & F1-S (\%) \\ \midrule
\multirow{9}{*}{Labelled P-MART} & \textit{Abstract Factory} & 48.8 & 52.3 & 50.49 & 73.33 & 71.15 & 72.22 & 78.33 & 78.33 & 78.33 \\
 & \textit{Adapter} & 15 & 20 & 17.14 & 78.14 & 75.62 & 76.86 & 91.6 & 86.66 & 89.06 \\
 & \textit{Builder} & 55 & 45 & 49.5 & 53.45 & 48.8 & 51.02 & 77.5 & 80 & 78.73 \\
 & \textit{Decorator} & 13.22 & 12.8 & 13.01 & 54.18 & 66 & 59.51 & 60 & 36.66 & 45.51 \\
 & \textit{Factory Method} & 50.23 & 40.35 & 44.75 & 78.23 & 80.1 & 79.15 & 56.67 & 63.33 & 59.82 \\
 & \textit{Observer} & 46.12 & 44.12 & 45.1 & 57.21 & 55.23 & 56.20 & 67.5 & 76.66 & 71.79 \\
 & \textit{Singleton} & 63 & 59 & 60.93 & 74.23 & 70.18 & 72.15 & 43.33 & 40.00 & 41.6 \\
 & \textit{Visitor} & 30.3 & 35.3 & 32.61 & 45.74 & 50.25 & 47.89 & 96 & 93.3 & 94.63 \\
 & \textit{None} & 57.23 & 70.02 & 62.98 & 51.3 & 51.67 & 51.48 & 78.5 & 82.64 & 80.52 \\ \cline{2-11}
  & \textit{\textbf{Overall}} & 42.1 & 42.1 & 41.83 & 62.87 & 63.22 & 63.04 & \colorbox{gray}{72.16} & \colorbox{gray}{70.84} & \colorbox{gray}{71.49} \\ \midrule
\multirow{9}{*}{DPD$_F$-Corpus} & \textit{Abstract Factory} & 55.5 & 49.5 & 52.33 & 75.5 & 77 & 76.24 & 93.27 & 92.08 & 92.67 \\
 & \textit{Adapter} & 35 & 31.5 & 33.16 & 85.16 & 78.25 & 81.56 & 71.56 & 66.55 & 68.96 \\
 & \textit{Builder} & 62.2 & 60.1 & 61.13 & 58.52 & 51.23 & 54.63 & 83.21 & 83.66 & 83.43 \\
 & \textit{Decorator} & 21.28 & 24.5 & 22.78 & 60.15 & 58.23 & 59.17 & 80.99 & 75 & 77.88 \\
 & \textit{Factory Method} & 61.3 & 50.45 & 55.35 & 82.15 & 80.8 & 81.47 & 73.58 & 81.88 & 77.51 \\
 & \textit{Observer} & 50.1 & 47.65 & 48.84 & 53.25 & 48.26 & 50.63 & 81.26 & 90 & 85.41 \\
 & \textit{Singleton} & 65 & 67 & 65.98 & 74.24 & 69.23 & 71.65 & 81.6 & 68.22 & 74.31 \\
 & \textit{Visitor} & 55.1 & 80.1 & 65.29 & 60.1 & 66.25 & 63.03 & 97.07 & 91.88 & 93.93 \\
 & \textit{None} & 65.25 & 79 & 71.47 & 51.3 & 56.24 & 53.67 & 70.44 & 75 & 72.65 \\ 
\cline{2-11}
 & \textit{\textbf{Overall}} & 52.30 & 54.42 & 52.93 & 66.71 & 65.05 & 65.87 & \colorbox{gray}{81.44} & \colorbox{gray}{80.47} & \colorbox{gray}{80.75} \\ \bottomrule
\end{tabular}
\label{tab:comp}
\end{table*}

\textbf{Comparison Strategy:} We apply the following strategy to compare our approach with state-of-the-art.

\begin{enumerate}
	\item We apply our approach on the corpus provided by the studies and compare results. This corpus is referred to as a \textit{P-MART Corpus}.
	\item  We apply selected studies to our labelled corpus \textit{DPD$_F$-Corpus} and compare results.
\end{enumerate}

\textbf{Labelling the P-MART Corpus:} We selected in total 290 files (including files not containing design patterns i.e none) from the P-MART corpus and labelled them using the same approach we use for labelling the DPD$_F$-Corpus. We refer to this corpus as Labelled P-MART Corpus. Table~\ref{tab:PMART-Label} lists the instances of design patterns used in the labelled P-MART corpus. The red coloured patterns are not identified by the classifier.

\begin{table}
\centering
\caption{Labelled instances of the benchmark corpus trained by the benchmark classifiers. The red coloured instances are not identified by the DPD$_F$ classifier.}
\begin{tabular*}{\tblwidth}{@{}LLLL@{}}  
\toprule
\textbf{Patterns} & \textbf{\#} & \textbf{Patterns} & \textbf{\#} \\ 
\midrule
Abstract Factory & 30    & Observer & 30  \\
Adapter          & 30    & \textcolor{red}{Prototype} & 26 \\
Builder          & 30    & \textcolor{red}{Proxy}     & \textcolor{red}{0}   \\
Decorator        & 23    & Singleton      & 12  \\
Factory Method   & 30    & Visitor        & 30   \\
\textcolor{red}{Fa\c{c}ade}  & \textcolor{red}{9}  & None & 30 \\
\textcolor{red}{Memento} & \textcolor{red}{10} & &  \\
\bottomrule
\end{tabular*}
\label{tab:PMART-Label}
\end{table}

\subsubsection*{Discussion}\label{subsubsec:_disc}
Here we compare our study and results with the state-of-the-art studies.

\textit{FeatureMaps:} \cite{Thaller:2019} apply feature-maps as an input to random forest and convolutional neural networks~\citep{Lecun:1998} to determine whether a given set of classes implement a particular pattern role. Their features are high-level conceptual classes such as micro-structures. As DPD$_F$ classifier's base estimator is Random Forest (RF), we only replicated and compared results with their Random Forest classifier. Though they have only identified \textit{Decorator} and \textit{Singleton} patterns, we replicate their study with all patterns labelled in the benchmark corpus i.e. creating feature maps for all patterns in the benchmark corpus. As shown in Table~\ref{tab:comp} our DPD$_F$ classifier outperforms the RF classifier in FeatureMaps in terms of Precision and Recall with approximately 30\%  improvements on benchmark corpus. With the DPD$_F$ corpus our study prevailed over theirs with approximately 30\% improvement in terms of Precision and 26\% in terms of Recall respectively.

\textit{MARPLE-DPD:} In this study,~\citep{Fontana:2011} have applied basic elements and metrics to mechanically extract design patterns from the source code and further extended it in~\citep{Zanoni:2015}. These approaches use different classifiers for the identification of design patterns, thus, we replicate their study only with the RF classifier and compare results with our DPD$_F$ classifier. We calculate the true and false positives for each pattern and calculate the weighted Precision and Recall for both classifiers. Results in Table~\ref{tab:comp} show that overall, our approach works far better than MARPLE-DPD with approximately 10\% improvements in terms of Precision on benchmark corpus and 15\% on DPD$_F$ corpus respectively.

To ensure fair predictions of labelled instances of design patterns by the classifiers, we make sure that the labelled benchmark corpus should have at least 30 instances for each pattern, which resulted in (total) seven patterns identified in the end from the benchmark corpus. The Fa\c{c}ade, Memento and Prototype are missed because the labelled instances (after removing duplicates) are 10 (as K=10 in cross validation) or less and can not be cross validated while training the classifiers. In the P-MART dataset many instances are labelled as decorators and prototypes so we treat them as decorator in the final labelled benchmark corpus. That is why there is no instance of Prototype selected for the labelled dataset. Both MARPLE and Featuremap identified singleton better than DPD$_F$ on the labelled P-MART Corpus. It is because very few instances of singleton are found in the corpus and thus not fully trained by the DPD$_F$ classifier. We have also generated results by setting the threshold value to 20 and the results were slightly poorer. From this we infer that the larger the dataset is, better the classifier will be trained and results with higher accuracy will be achieved.

In summary, FeatureMaps works poorly on both corpora whereas, MARPLE-DPD identifies most of the patterns reasonably well with over 62\% Precision on benchmark corpus and approximately 67\% on DPD$_F$ corpus. Nevertheless, it still lags behind our approach by approximately 10\% on benchmark corpus and approximately 15\% on our corpus in terms of Precision.

\section{Threats to validity}\label{sec:threats}
In the following section, we outline the relevant threats to the validity of our work.

\subsection{Threats to internal validity}\label{subsec:in_threats}
\textbf{Counting of instances:} The first internal threat to the validity of our study is the minor difference in the counting methodologies used between our and benchmark studies. The benchmark studies counted patterns using pattern roles whereas we counted the instances of pattern contained in the file. In addition the DPExample project was not publicly available, which may slightly change the results. We alleviate this problem by labelling the benchmark corpus using our method and compare results.

\textbf{Bugs:} It is possible that there are bugs in the processing code that affect the SSLR file generation and the training data, as well as bugs in the \textit{Word2Vec} model and the classifiers we have used, though this is less likely to be a problem given the wide use these tools have already received. We have spent some time debugging our code, found and fixed minor issues, which tend to improve the classification accuracy, so it is more likely than not that bugs will lead to lower precision and/or Recall than would otherwise be the case.

\textbf{Data Labelling:} Though we have labelled data using an online tool, there is a possibility of disagreement between the labellers, which may lead to incorrect labelling of the data. To mitigate these issues, we hire at least three labellers to label the corpus as discussed in section~\ref{subsubsec:data_col}. This process has substantially reduced the disagreement among raters as shown by the high kappa score. To further reduce the disagreement, we intend to hire more labellers in future.

\subsection{Threats to external validity}\label{subsec:ex_threats}
The reference set may not reflect the totality of Java source code that may be of interest to developers, and it is unknown whether the classifier will work effectively on the source code of interest outside the reference set. Obviously, the best way to mitigate this risk is to further increase the size and diversity of the reference set, which we plan to address in future.

\section{Related Work}\label{sec:relatedwork}
During the past years, with the growing amount of electronically available information, there is substantial interest and a substantial body of work among software engineers and academic researchers in design pattern detection. A majority of the approaches to the detection of design patterns transform the source code and design patterns into some intermediate representations such as rules, models, graphs, productions and languages~\citep{Yu:2018}. For example, \citep{Bernardi:2013} exploited a meta-model which contains a set of properties characterising the structures and behaviours of the source code and design patterns and a matching algorithm is performed to identify the implemented patterns. \citep{Alnusair:2014} employed semantic rules to capture the structures and behaviours of the design patterns, based on which the hidden design patterns in open-source libraries are discovered. Recently, \citep{Xiong:2019} applied ontology based parser with idiomatic phrases to identify design patterns achieving high Precision.

A good number of studies developed tools that used source code or its intermediate representation to identify patterns as well as machine learning models to predict patterns.~\citep{De:2011},~\citep{Tosi:2009} and~\citep{Zhang:2013} used static and dynamic analysis (or a combination of both) to develop Eclipse plugins that detect design patterns from source code.~\citep{Moreno:2012} developed a tool \textit{JStereoType} used for detecting low-level patterns (classes, interfaces etc) to find the design intent of the source code.~\citep{Zanoni:2015} exploited a combination of graph matching and machine learning techniques to implement a tool called MARPLE-DPD. ~\citep{Niere:2002} designed the FUJABA Tool Suite which provides developers with support for detecting design patterns (including their variants) and smells.~\citep{Hautamaki:2005} used a pattern based solution and tool to teach software developers how to use development solutions in a project. 

Some techniques (including tool generation) applied reverse engineering to identify design patterns from source code and UML artefacts. For instance,~\citep{De:2011} built a tool using static analysis and applied reverse engineering through visual parsing of diagrams. Other studies such as~\citep{Thongrak:2014, Panich:2016, Shi:2006} also applied reverse engineering techniques on UML class and sequence diagrams to extract design patterns using ontology. \citep{Brown:1996} proposed a method for automatic detection of design patterns by reverse-engineering the SmallTalk code.

Very few studies utilised code metrics in identification of design patterns.~\citep{Uchiyama:2011} have presented a software pattern detection approach by using software metrics and machine learning techniques. They have identified candidates for the roles that compose the design patterns by considering machine learning and software metrics. \citep{Lanza:2007} approach uses learning from the information extracted from design pattern instances which normally include variant implementations such as number of accessor methods etc.~\citep{Fontana:2011} introduced the micro-structures that are regarded as the building blocks of design patterns.~\citep{Thaller:2019} has built a feature map for pattern instances using neural networks.

Several other approaches exploit machine learning to solve the issue of variants. For example,~\citep{Chihada:2015} mapped the design pattern detection problem into a learning problem. Their proposed detector is based on learning from the information extracted from design pattern instances, which normally include variant implementations. ~\citep{Ferenc:2005} applied machine learning algorithms to filter false positives out of the results of a graph matching phase, thus providing better precision in the overall output while considering variants. A recent work by \citep{Hussain:2018} leverage deep learning algorithms for the organisation and selection of DPs based on text categorisation. To reduce the size of training examples for DP detection, a clustering algorithm is proposed by \citep{Dong:2008} based on decision tree learning. 

Though we have used machine learning based classifiers to test the efficacy of our approach, our approach is substantially different from the aforementioned studies as well as the benchmark studies in many ways. We have summarised the differences as under:

\begin{itemize}
    \item We have identified and employed 15 source code features whereas as the benchmark studies of~\citep{Zanoni:2015} utilised code metrics and \citep{Thaller:2019} used feature maps.
    \item Our labelled corpus size is large having 1,300 files extracted from more than 200 projects from the GJC. \citep{Zanoni:2015} and \citep{Thaller:2019} used a P-MART corpus that has 1039 files containing the design patterns we intend to identify in a highly imbalanced nature of the P-MART corpus.
    \item Our machine learning classifier achieved approximately 80\% precision whereas existing studies achieved 42\% and 63\% Precision on Labelled P-MART corpus and 52\% and 67\% on the DPD$_F$ corpus respectively.
    \item Our classifier identified 12 design patterns successfully, whereas existing studies only recognised two and six design patterns respectively.
\end{itemize}

\section{Conclusion \& Future Work}\label{sec:conclusion}
In this paper, we introduce DPD$_F$, which is a novel approach for detecting software design patterns by using source code features and machine learning methods. DPD$_F$ builds an SSLR representation by applying call graph and source code features on the DPD$_F$ corpus extracted from the publicly available \textit{'The Java Github Corpus'}. Next, DPD$_F$ constructs a Java n-gram model by applying the \textit{Word2Vec} algorithm on the SSLR file. Finally, DPD$_F$ trains a supervised machine learning classifier on the labelled DPD$_F$ corpus to detect design patterns. 

To statistically evaluate the efficacy of the proposed approach we apply three commonly used statistical measures namely Precision, Recall and F1-Score, and build a confusion matrix to verify the efficacy of our classifier. Empirical results show that our proposed approach DFD$_F$ detects software design patterns with approximately 80\% on Precision and 79\% on Recall. DPD$_F$ outperforms two existing approaches, by 30\% and 15\% respectively in terms of Precision, as well as is able to detect more patterns (12) compared to existing studies.

In future, we plan to investigate whether other useful information for software maintainers can be extracted using our code feature or by other more features. While our classifier's performance is promising, further improvements are clearly highly desirable for practical use. While increasing the size of the training set will almost certainly improve accuracy, it is likely that adding additional code features to the SSLR files would also help, and could be targeted to improve performance on those patterns where accuracy is relatively low.

\appendix
\section{Reproducibility}
For the purpose of reproducibility of our results, we have released our complete implementation with the annotated reference set and the result files to the public as an open-source project via our online appendix at~\url{https://github.com/najamnazar/designpatterndetection}.
The codelabeller online tool's code is accessible at \url{https://github.com/najamnazar/codelabeller}.

\section*{Acknowledgement}
We would like to extend our gratitude to the individuals who have dedicated their time and effort in performing annotation of design patterns through the online tool. We would also like to thank all anonymous reviewers for their valuable comments. This research received no internal or external funding from any public or private agency.
\bibliographystyle{cas-model2-names}
\bibliography{paper}

\end{document}